\documentclass[fleqn,12pt,twoside]{article}
\usepackage{espcrc1}

\usepackage{epsfig}

\usepackage{floatflt}

\newcommand\sqrtsnn{\sqrt{s_{_{NN}}}}
\newcommand\pT{p_{_{T}}}
\newcommand\AuAu{${\rm Au} + {\rm Au}$}
\newcommand\PbPb{${\rm Pb} + {\rm Pb}$}
\newcommand\ee{${\rm e}^+ {\rm e}^-$}
\newcommand\pp{${\rm p} {\rm p}$}
\newcommand\pA{${\rm p} {\rm A}$}
\newcommand\ppbar{${\rm p} \bar{\rm p}$}
\newcommand\avgNp{\langle N_{part} \rangle}
\newcommand\hlfNp{\langle N_{part}/2 \rangle}

\title{Global Observations from PHOBOS}
\author{Mark D.~Baker for the PHOBOS Collaboration}

\begin{document}

\maketitle
\noindent
{\small
B.B.Back$^1$,
M.D.Baker$^2$,
D.S.Barton$^2$,
R.R.Betts$^6$,
M.Ballintijn$^4$,
A.A.Bickley$^7$,
R.Bindel$^7$,
A.Budzanowski$^3$,
W.Busza$^4$,
A.Carroll$^2$,
M.P.Decowski$^4$,
E.Garc\'{\i}a$^6$,
N.George$^{2}$,
K.Gulbrandsen$^4$,
S.Gushue$^2$,
C.Halliwell$^6$,
J.Hamblen$^8$,
G.A.Heintzelman$^2$,
C.Henderson$^4$,
D.J.Hofman$^6$,
R.S.Hollis$^6$,
R.Ho\l y\'{n}ski$^3$,
B.Holzman$^2$,
A.Iordanova$^6$,
E.Johnson$^8$,
J.L.Kane$^4$,
J.Katzy$^{4,6}$,
N.Khan$^8$,
W.Kucewicz$^6$,
P.Kulinich$^4$,
C.M.Kuo$^5$,
W.T.Lin$^5$,
S.Manly$^8$,
D.McLeod$^6$,
J.Micha\l owski$^3$,
A.C.Mignerey$^7$,
R.Nouicer$^6$,
A.Olszewski$^3$,
R.Pak$^2$,
I.C.Park$^8$,
H.Pernegger$^4$,
C.Reed$^4$,
L.P.Remsberg$^2$,
M.Reuter$^6$,
C.Roland$^4$,
G.Roland$^4$,
L.Rosenberg$^4$,
J.Sagerer$^6$,
P.Sarin$^4$,
P.Sawicki$^3$,
W.Skulski$^8$,
S.G.Steadman$^4$,
P.Steinberg$^{2,b}$,
G.S.F.Stephans$^4$,
M.Stodulski$^3$,
A.Sukhanov$^2$,
J.-L.Tang$^{5,a}$,
R.Teng$^8$,
A.Trzupek$^3$,
C.Vale$^4$,
G.J.van~Nieuwenhuizen$^4$,
R.Verdier$^4$,
B.Wadsworth$^4$,
F.L.H.Wolfs$^8$,
B.Wosiek$^3$,
K.Wo\'{z}niak$^3$,
A.H.Wuosmaa$^{1,c}$,
B.Wys\l ouch$^4$\\
\vspace{2mm}
\small

\noindent
$^1$~Argonne National Laboratory, Argonne, IL 60439-4843, USA\\
$^2$~Brookhaven National Laboratory, Upton, NY 11973-5000, USA\\
$^3$~Institute of Nuclear Physics, Krak\'{o}w, Poland\\
$^4$~Massachusetts Institute of Technology, Cambridge, MA 02139-4307, USA\\
$^5$~National Central University, Chung-Li, Taiwan\\
$^6$~University of Illinois at Chicago, Chicago, IL 60607-7059, USA\\
$^7$~University of Maryland, College Park, MD 20742, USA\\
$^8$~University of Rochester, Rochester, NY 14627, USA\\
$^a$~Current address: National Chung-Cheng University, Chia-Yi, Taiwan\\
$^b$~Current address: University of Cape Town, South Africa\\
$^c$~Current address: Western Michigan University, Kalamazoo, MI 49008, USA\\
}

\begin{abstract}
Particle production in Au+Au collisions has been measured in the
PHOBOS experiment at RHIC for a range of collision energies.  Three
empirical observations have emerged from this dataset which require
theoretical examination. First, there is clear evidence of limiting
fragmentation. Namely, particle production in central \AuAu\
collisions, when expressed as $dN/d\eta'$ ($\eta' \equiv
\eta-y_{beam}$), becomes energy independent at high energy for a broad
region of $\eta'$ around $\eta'=0$.  This energy-independent region
grows with energy, allowing only a limited region (if any) of
longitudinal boost-invariance. Second, there is a striking similarity
between particle production in \ee\ and \AuAu\ collisions (scaled by
the number of participating nucleon pairs).  Both the total number of
produced particles and the longitudinal distribution of produced
particles are approximately the same in \ee\ and in scaled \AuAu. This
observation was not predicted and has not been explained.  Finally,
particle production has been found to scale approximately with the
number of participating nucleon pairs for $\avgNp>65$. This scaling
occurs both for the total multiplicity and for high $\pT$ particles
(3~$<\pT<$~4.5~GeV/c).
\end{abstract}

\section{INTRODUCTION}

\begin{figure}[htbp]
\centerline{
\epsfig{file=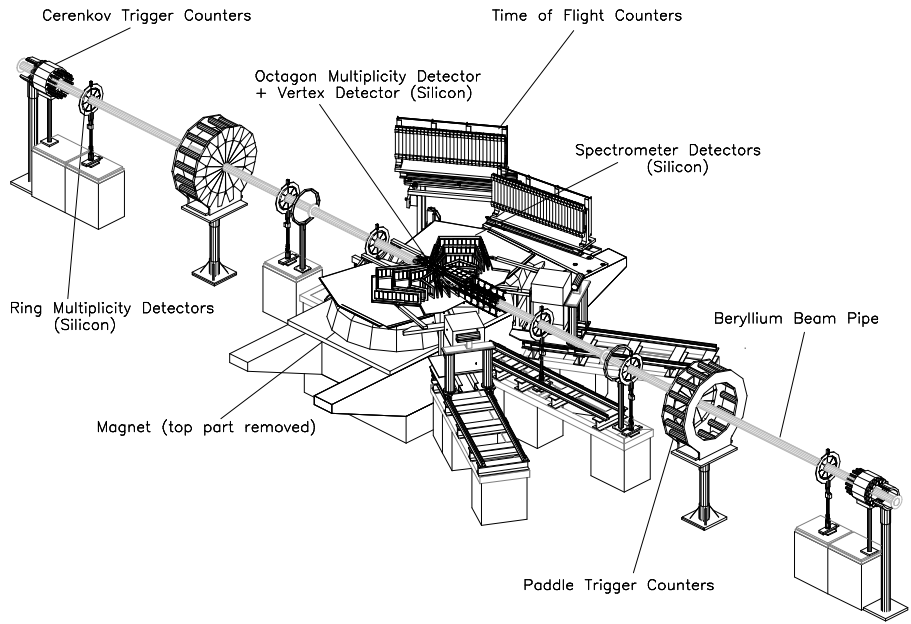,width=12cm}}
\vspace*{-0.5cm}
\caption{Schematic diagram of the detector setup for the 2001 running
period. }
\label{fig:detector}
\end{figure}

The data described in this paper were taken during the Year 2000 and
Year 2001 runs of the Relativistic Heavy Ion Collider at Brookhaven
National Laboratory. The PHOBOS apparatus~\cite{phobos2}, shown
schematically in Figure~\ref{fig:detector}, was used to take data at
three energies: $\sqrtsnn=19.6$, 130, and 200~GeV. A partial test
detector was also used to take data at $\sqrtsnn=56$~GeV~\cite{phobosprl}.

The full PHOBOS apparatus comprises several subdetectors. Silicon
detectors are employed for vertex finding, particle tracking and
multiplicity measurements.  This set of detectors has nearly full
azimuthal coverage over a large pseudorapidity range $|\eta| < 5.4$.
The detector setup also includes two sets of 16 scintillator counters
(``paddle counters'') located at $\pm3.21$~m relative to the nominal
interaction point along the beam ($z$) axis. These counters cover
pseudorapidity in the range $3 < |\eta |< 4.5$ and served as the
primary event trigger.  The collision centrality is characterized by
the average number of nucleon participants~$\avgNp$, determined as
described in Refs.~\cite{phoboscent200,limfrag}. These references also
tabulate the values and systematic errors of $\avgNp$ for each
centrality bin.

\section{SURVEY OF BASIC MEASUREMENTS}

A key goal in the design of the PHOBOS experiment was to perform a
broad and systematic survey of hadronic particle production in heavy
ion collisions.  While this survey is still incomplete in terms of
energies and species explored, significant progress has been made
toward this goal.  This section summarizes the basic measurements
available so far: multiplicity, particle spectra and azimuthal
asymmetry of particle production.

The pseudorapidity densities ($dN_{ch}/d\eta$) and particle yields
($d^2N/dyd\pT$) given here refer to primary produced particles and do
not include feed-down products from weak decays of neutral strange
particles.  Corrections were made for residual effects from secondary
interactions and weak decay feed-down as well as for particles which
were absorbed or produced in the material surrounding the collision
(primarily the Beryllium beampipe and the magnet steel).

\begin{figure}[htbp]
\centerline{ \epsfig{file=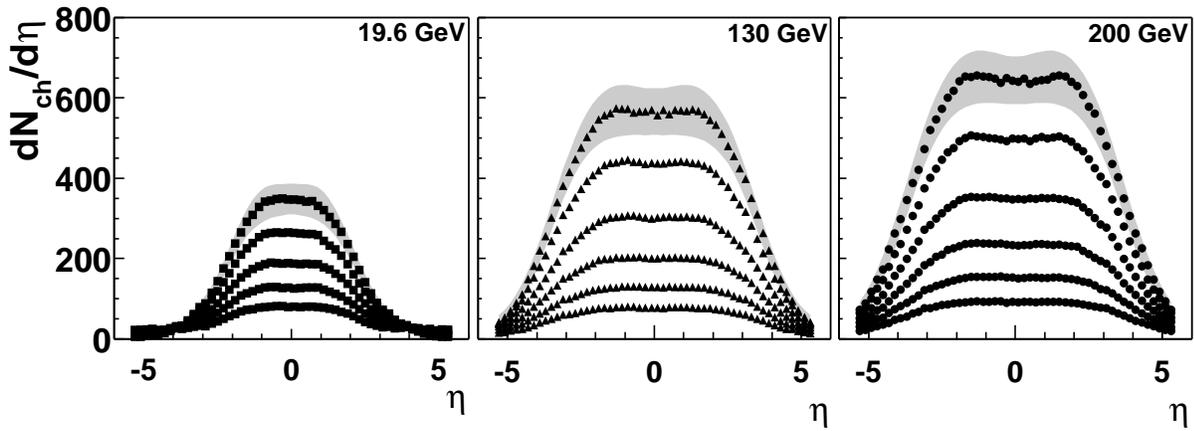,width=16cm}}
\vspace*{-0.5cm}
\caption{Charged particle pseudorapidity distribution,
$dN_{ch}/d\eta$, measured for \AuAu\ at $\sqrt{s_{_{NN}}} =$~19.6,
130, and 200~GeV for the centrality bins: 0--6\%, 6--15\%, 15--25\%,
25--35\%, 35--45\%, 45--55\%. The most peripheral bin is excluded for
the lowest energy. The statistical error is negligible. The typical
systematic error is shown as a 90\% C.L.\ band for selected centrality
bins.}
\label{fig:dNdetas}
\end{figure}

The unique features of the PHOBOS experiment relevant for the charged
particle multiplicity measurements are the large acceptance of the
detector and the 12~m long, 1~mm thick, Beryllium beampipe which
limits the absorption of low momentum particles and the production of
secondaries.  Figure~\ref{fig:dNdetas} shows the charged particle
pseudorapidity distributions ($dN_{ch}/d\eta$) measured at
$\sqrt{s_{_{NN}}} =$~19.6, 130, and 200 GeV for a variety of
centrality bins for pseudorapidity in the range
$-5.4<\eta<5.4$~\cite{limfrag}.  Due to the large coverage in $\eta$,
$dN_{ch}/d\eta$ is measured over almost the full range, except for a
small missing fraction at very high $|\eta|$. For central events, this
missing fraction is estimated to be less than 2\%. The multiplicity
results are discussed further in Sections~\ref{sec:rule1}
and~\ref{sec:rule2}.

\begin{figure}[htb]
\centerline{ \epsfig{file=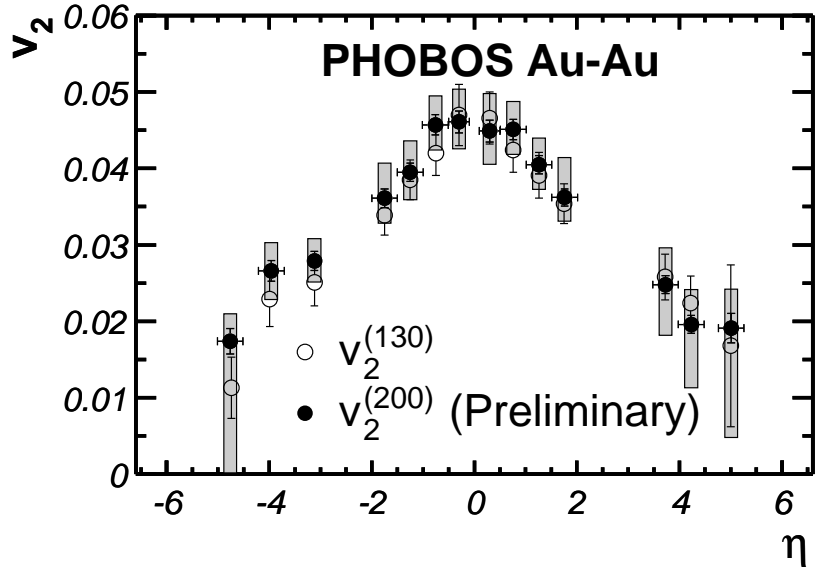,width=12cm} }
\vspace*{-0.5cm}
\caption{Elliptic flow as a function of pseudorapidity for Au-Au
collisions at $\sqrt{s_{_{NN}}}$ = 130 and 200 GeV.}
\label{fig:v2etaener}
\end{figure}

Figure~\ref{fig:v2etaener} shows the results of elliptic flow as a
function of pseudorapidity from PHOBOS for 130 and 200 GeV.  A
unique feature of these results is the broad pseudorapidity reach
covering almost 11 units of pseudorapidity and showing that there is
no broad boost-invariant region near mid-rapidity. The error bars
represent 1$\sigma$ statistical errors and the boxes represent 90\%
confidence level systematic errors for the 200 GeV data points.  These
results were measured using a hit-based analysis~\cite{flowprl}.  A
new, track-based, analysis is also available both as a cross-check and
to provide $v_2$ as a function of particle $\pT$~\cite{flowQM02}.

\begin{figure}[htb]
\vspace*{-0.5cm}
\begin{minipage}[t]{75mm}
\epsfig{file=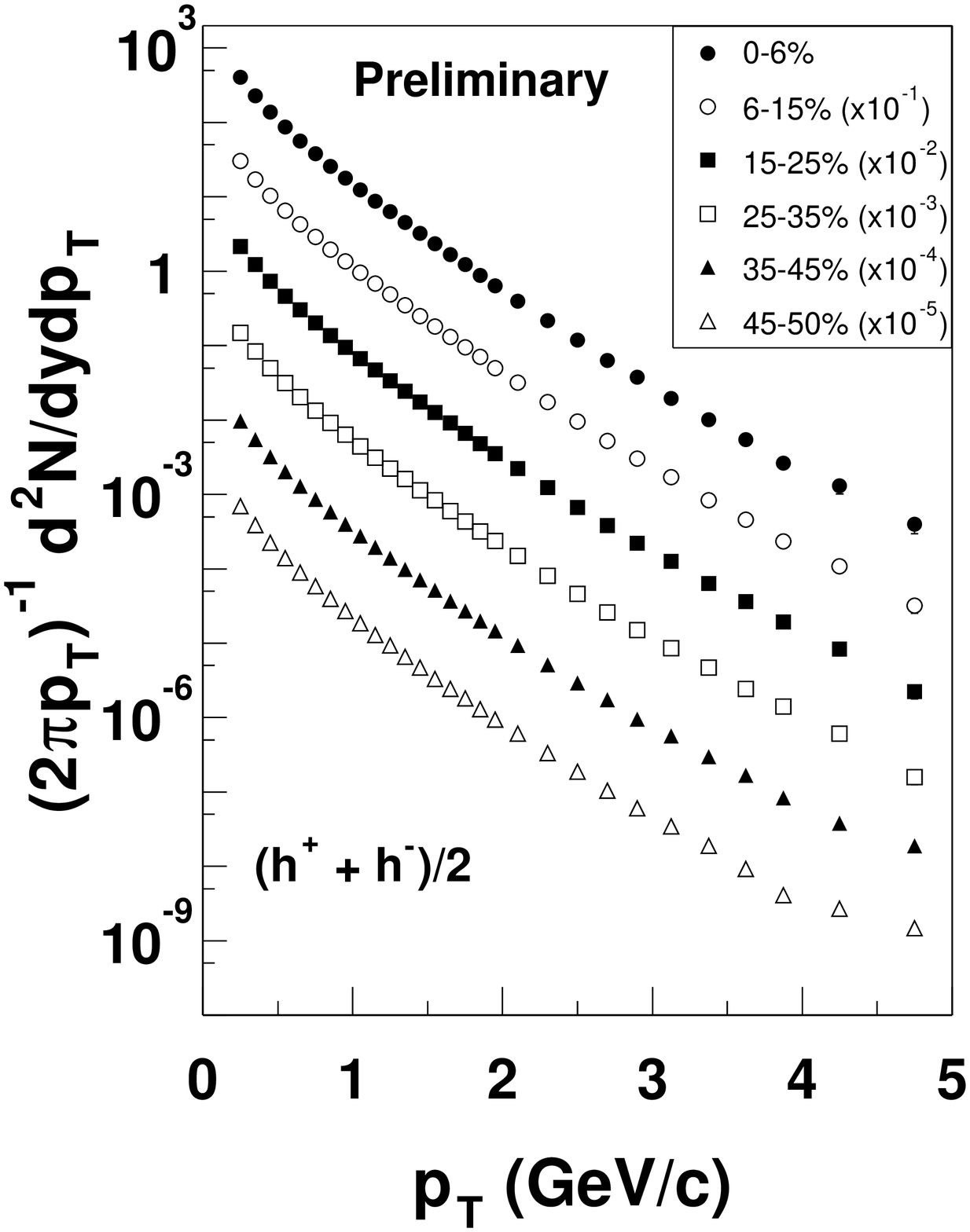,width=69mm}
\vspace*{-0.5cm}
\caption{Charged particle invariant yields in \AuAu\ collisions for a
variety of centrality ranges. Non-central data have been scaled by
arbitrary factors labeled on the plot for clarity.}
\label{fig:spectra}
\end{minipage}
\hspace{\fill}
\begin{minipage}[t]{75mm}
\epsfig{file=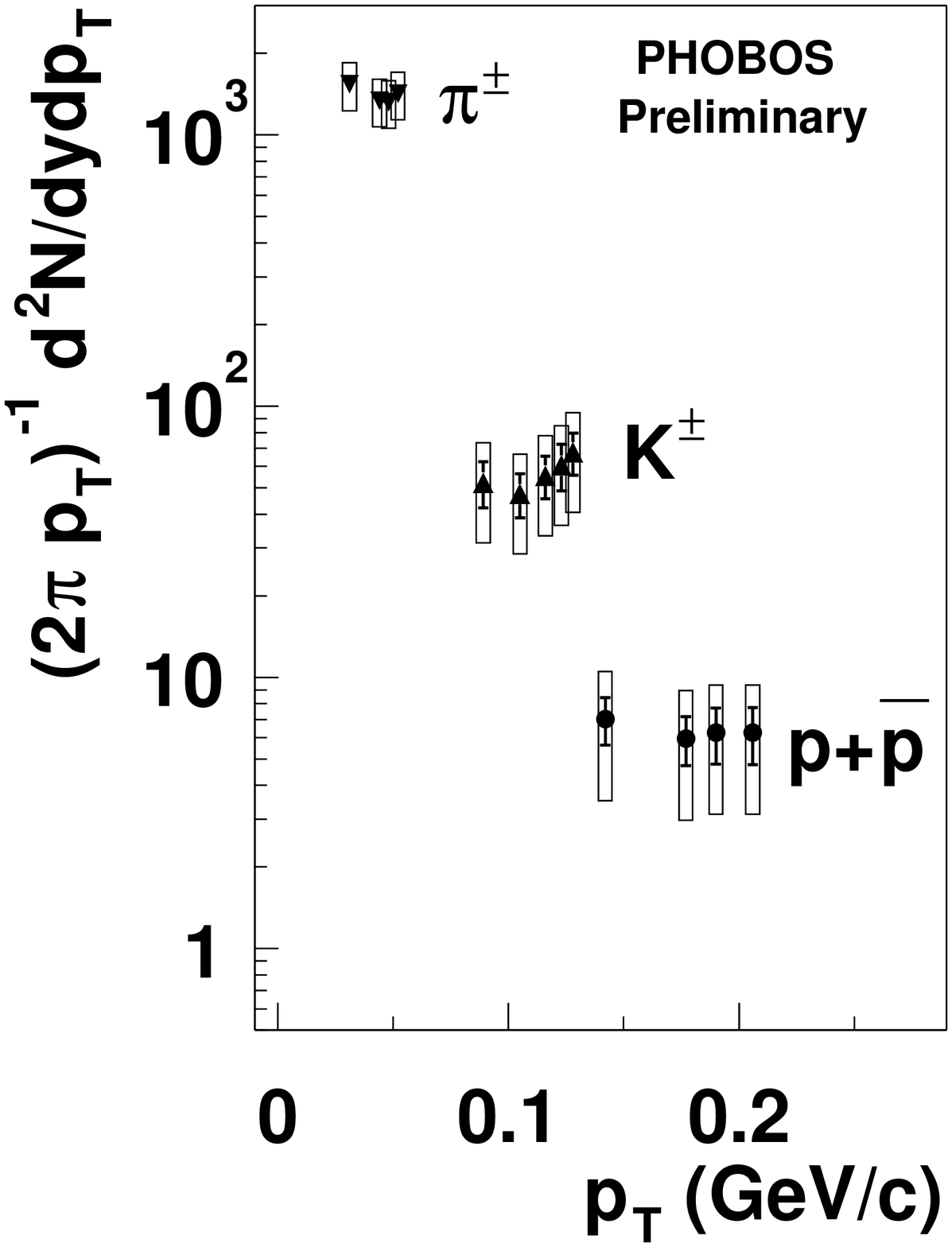,width=73mm}
\vspace*{-0.5cm}
\caption{Identified particle invariant yields as a function of $p_T$ for
central (0--15\%) \AuAu\ collisions. 
The boxes show systematic uncertainties.}
\label{fig:lowpt}
\end{minipage}
\end{figure}

Figure~\ref{fig:spectra} shows the unidentified charged particle yield
measured by PHOBOS for a range of centralities, with rapidity,
$y_\pi$, defined using the pion mass.  These data cover rapidity in
the range $0.2 < y_\pi < 1.4$ and transverse momenta in the range $0.2
< \pT < 5.0$~GeV/c, corresponding to six decades of change in the
magnitude of the yield. Figure~\ref{fig:lowpt} shows the invariant
yields at very low $\pT$ measured by PHOBOS~\cite{BWQM02}. These
particles, which make it through the beampipe and then range out in
the active Silicon detectors in the field free region of the
spectrometer, are mass-identified but not charge-identified.  The
yields are shown for ($\pi^{\pm}$), ($K^{\pm}$) and ($p +
\overline{p}$) measured at mid-rapidity $(-0.1<y<0.4)$ in the
transverse momentum ranges from 30 to 50 MeV/c for charged pions, 90
to 130 MeV/c for kaons and 140 to 210 MeV/c for protons and
antiprotons for the 15\% most central Au+Au collisions at
$\sqrt{s_{_{NN}}}$ = 200 GeV.

The unique features of PHOBOS relevant for
the momentum-measured particles are the close proximity of the
detector to the interaction region~(10--80~cm), the precise vertex
determination~(0.1--0.3~mm), good segmentation~(0.4--1~mm) of the silicon
detectors in the bend direction and, again, the small amount of
material between the interaction and the first layers of
silicon. These features allow the measurement of particles with good
momentum resolution over a broad range of transverse momenta from
0.03--5.0~GeV/c and beyond (when statistics allow), as well as
providing the ability to reject most secondaries and decay products.

The broad $\pT$ range, particularly the coverage to very low $\pT$,
will be important for constraining the role of dynamical processes
such as rescattering and radial expansion in these collisions. The
charged particle multiplicity (and elliptic flow) measurement spans
roughly 11 units of pseudorapidity, $2\pi$ in azimuth, and a factor of
10 in beam energy, in a single experiment, providing strong
constraints on any description of the initial state and subsequent
dynamics. From this broad dataset, three empirical scaling
rules have emerged. Sections~\ref{sec:rule1}--\ref{sec:rule3} describe
each of them in turn. These rules, at the very least, provide a
compact description of the data which must be respected by any
model. At best, these empirical scaling rules may point the way to a
more accurate description of the dynamics of these collisions.

\section{THE LIMITING CURVE FOR PARTICLE PRODUCTION}

\label{sec:rule1}

\begin{floatingfigure}[r]{8cm}
\vspace*{0.55em}
\epsfig{file=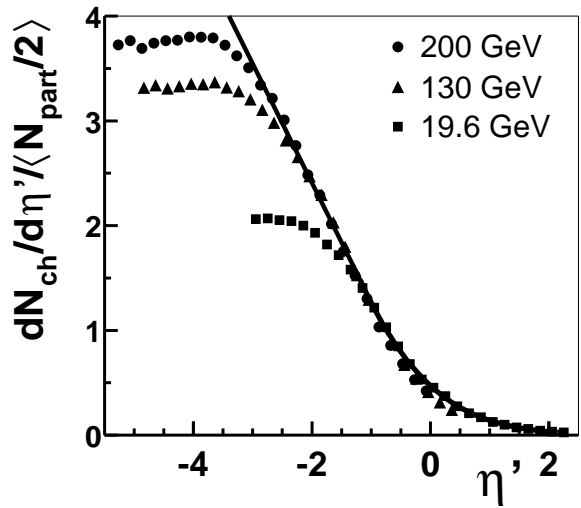,width=75mm}
\vspace*{-0.5cm}
\caption{Shifted pseudorapidity distribution, $dN_{ch}/d\eta'$ per
participant pair, where $\eta' \equiv \eta - y_{beam}$, for \AuAu\
data at $\sqrt{s_{_{NN}}} =$ 19.6, 130, and 200~GeV.  Systematic
errors not shown. The curve is to ``guide the eye''.}
\label{fig:limfragc}
\end{floatingfigure}

The pseudorapidity distributions were shown in
Figure~\ref{fig:dNdetas} for three different collision energies
($\sqrtsnn$).  In order to separate the trivial kinematic broadening
of the $dN_{ch}/d\eta$ distribution from the more interesting
dynamics, the data for \AuAu\ collisions at different energies can be
viewed in the rest frame of one of the colliding nuclei. Such an
approach led to the ansatz of ``limiting
fragmentation''~\cite{Yanglimfrag}, which successfully predicted the
energy dependence of particle production away from mid-rapidity in
hadron collisions, including pA~\cite{Elias80} and
\ppbar~\cite{UA5limfrag}. This ansatz states that, at high enough
collision energy, both \mbox{$d^2N/dy'dp_{_T}$} and the mix of
particle species (and therefore also $dN/d\eta'$) reach a limiting
value and become independent of energy in a region around $y'\sim 0$,
where \mbox{$y'\equiv y-y_{beam}$} and rapidity \mbox{$y \equiv
\tanh^{-1} \beta_z$}, with the $\hat{z}$ axis defined as the beam
(collision) axis.

Figure~\ref{fig:limfragc} shows the scaled, shifted pseudorapidity
distributions \mbox{$dN_{ch}/d\eta'/\langle N_{part}/2\rangle$}~\cite{limfrag}.
The results are folded about mid-rapidity (positive and negative
$\eta$ bins are averaged).  The distributions are observed to be
independent of collision energy over a substantial $\eta'$ range.
This is consistent with and extends a similar observation made by
BRAHMS~\cite{brahms200} over a more restricted $\eta'$ range.  Both
the 19.6 and 130~GeV data reach 85--90\% of their maximum value before
deviating significantly (more than 5\%) from the common limiting
curve.

The data presented here demonstrate that limiting fragmentation
applies in the \AuAu\ system, and that the ``fragmentation region'' in
\AuAu\ is rather broad, covering more than half of the available range
of $\eta'$ over which particles are produced. In particular, the
fragmentation region grows significantly between 19.6~GeV and 130 GeV,
extending more than two units away from the beam rapidity.  Particle
production appears to approach a fixed limiting curve which extends
far from the original beam rapidity, indicating that this limiting
curve is an important feature of the overall interaction and not
simply a nuclear breakup effect. This result is in sharp contrast to
the boost-invariance scenario~\cite{Bj} which predicts a fixed
fragmentation region and a broad central rapidity plateau that grows
in extent with increasing energy.

\medskip

\section{SIMILARITY OF AA AND $e^+e^-$ AT HIGH ENERGY}

\label{sec:rule2}

\begin{figure}[htb]
\begin{minipage}[t]{75mm}
\epsfig{file=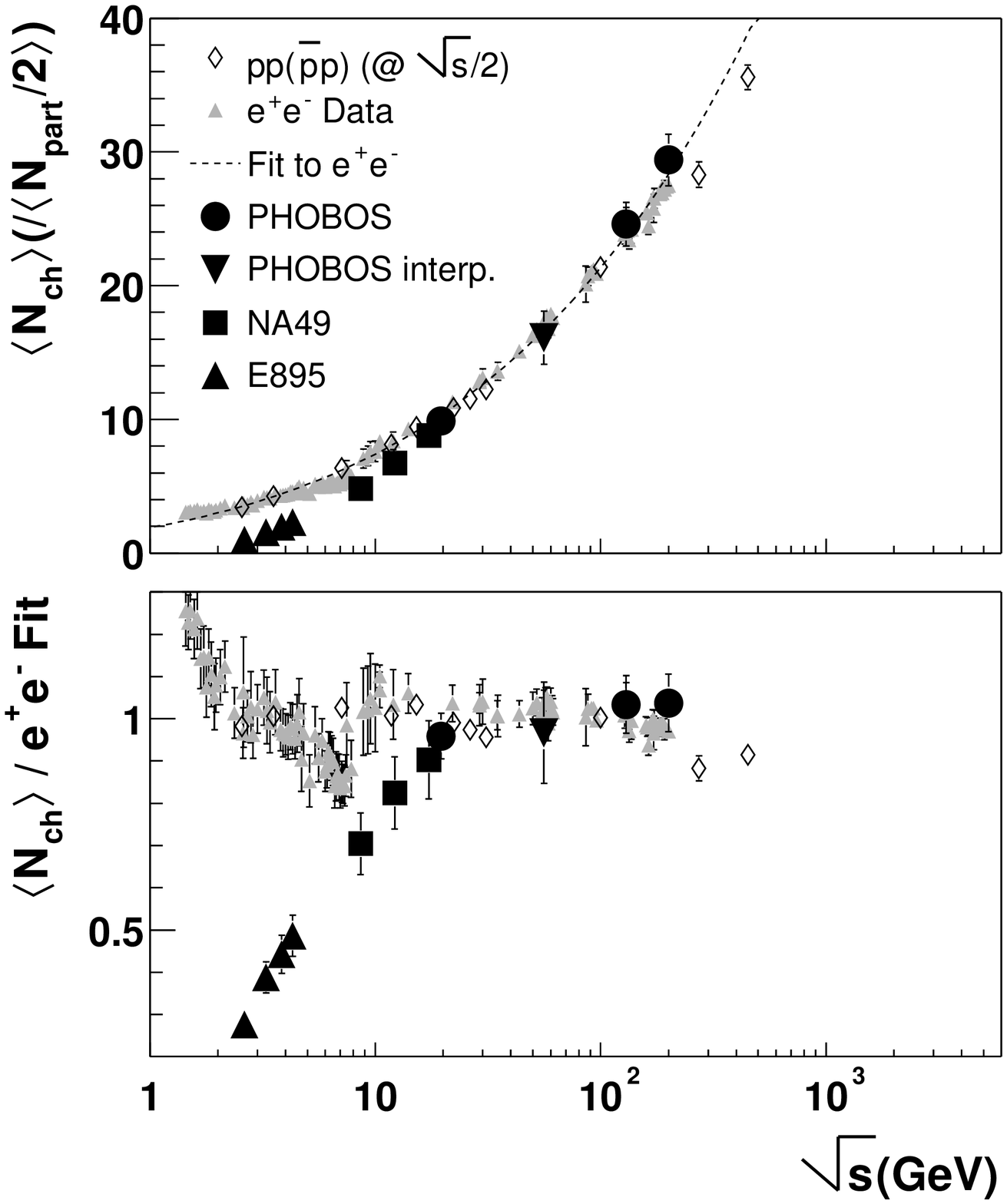,width=75mm}
\vspace*{-0.5cm}
\caption{Comparison of the total charged multiplicity versus collision
energies for AA, \ee, \pp, and \ppbar\ data, as described in the
text. In the upper panel, the curve is a perturbative QCD expression
fit to the \ee\ data. In the lower panel, the data have all been
divided by the \ee\ fit.}
\label{fig:AAee}
\end{minipage}
\hspace{\fill}
\begin{minipage}[t]{75mm}
\epsfig{file=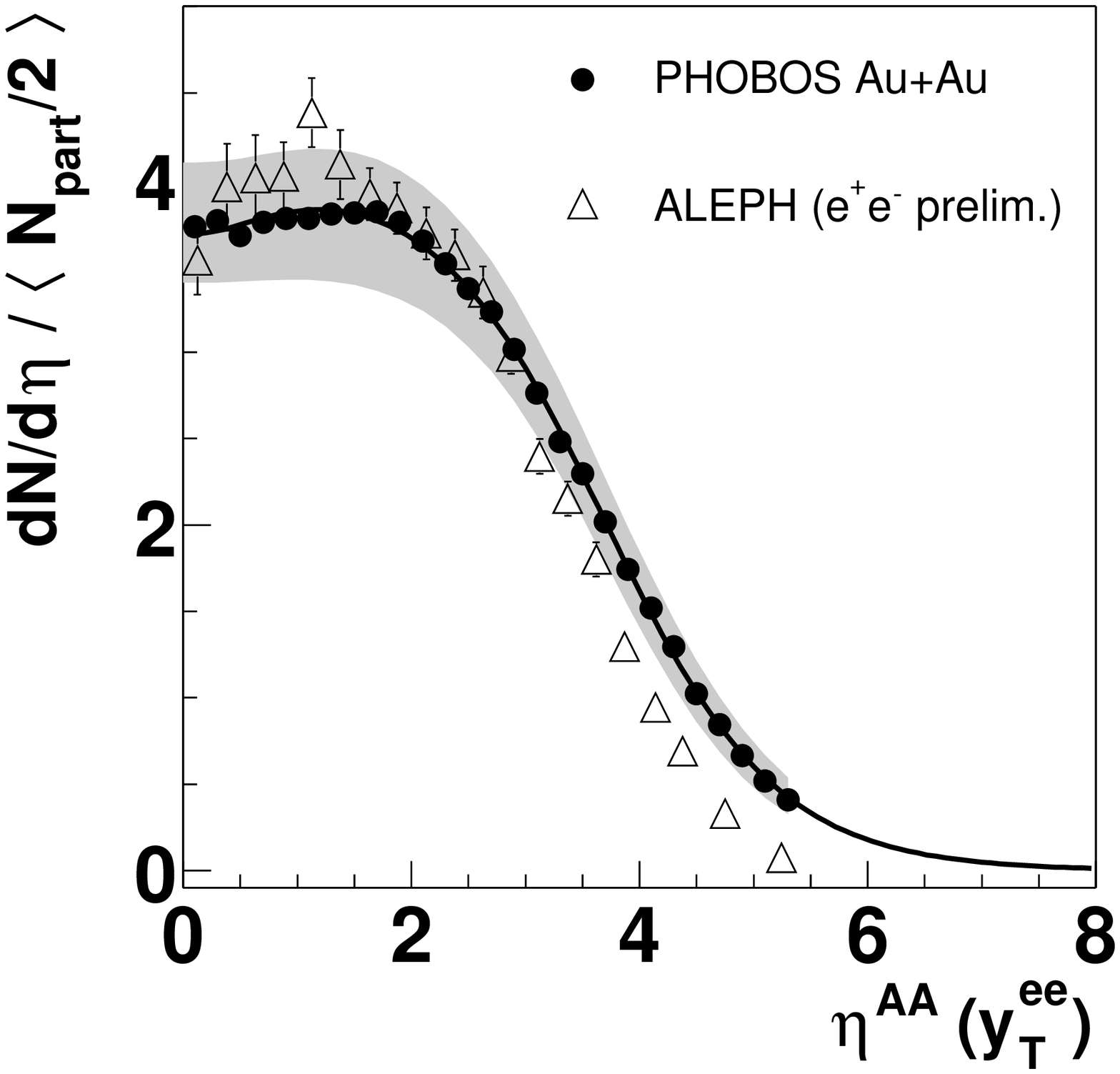,width=75mm}
\vspace*{-0.5cm}
\caption{Pseudorapidity distribution of charged particles produced
in central \AuAu\ collisions at $\sqrtsnn=200$~GeV compared with the
rapidity distribution along the thrust axis of particles produced in
\ee\ collisions at $\sqrt{s}=200$~GeV. The \AuAu\ data are normalized by
$N_{part}/2$. Systematic errors are shown for the \AuAu\ data.}
\label{fig:AAeeshape}
\end{minipage}
\end{figure}

In the upper part of Figure~\ref{fig:AAee}, total charged multiplicity
data from \pp, \ppbar, \ee, and AA (scaled by $\avgNp/2$) are shown as
a function of the appropriate $\sqrt{s}$ for each system. See
Ref.~\cite{PASQM02} for more details and
\cite{otherdata} for the non-PHOBOS data.  The \ee\ data
serves as a reference sample, describing the behavior of a simple
color dipole system with a large $\sqrt{s}$. The curve is a
description of the \ee\ data, given by the functional form: $C
\alpha_{s}(s)^A e^{\sqrt{B/\alpha_{s}(s)}}$ with the parameters A and
B calculable in perturbative QCD and the constant parameter C
determined by a fit to the \ee\ data~\cite{Mueller}. In order to
compare them with \ee, the \pp\ and \ppbar\ data were plotted at an
effective energy $\sqrt{s_{eff}}=\sqrt{s}/2$, which approximately
accounts for the ``leading particle effect''~\cite{Basile}. Finally,
central AA collisions, \AuAu\ from the AGS and RHIC, and \PbPb\ from
CERN are shown. Over the available range of RHIC energies from 19.6 to
200~GeV, the \AuAu\ results are consistent with the \ee\ results,
suggesting a universality of particle production at high energy. In
addition, the \AuAu\ data approximately agrees with the scaled \pp\
and \ppbar\ data suggesting that the effective energy of a high energy
AA collision is approximately just $\sqrtsnn$. The lower part of
Figure~\ref{fig:AAee} shows these same results, but divided by the
\ee\ fit. This figure illustrates the approach of AA from lower
energies towards the high energy scaling result.

Figure~\ref{fig:AAeeshape} compares the pseudorapidity distribution
for central \AuAu\ collisions to the closest available analog in \ee\
collisions: the thrust-axis-rapidity distribution, where the pion mass
was used for all charged particles in \ee~\cite{delphi}. This shows
that particle production in \AuAu\ and \ee\ collisions approximately
agree in longitudinal particle distribution as well as in the total
yield.

\section{PARTICIPANT SCALING OF PARTICLE PRODUCTION}

\label{sec:rule3}

\begin{figure}[htb]
\begin{minipage}[t]{75mm}
\epsfig{file=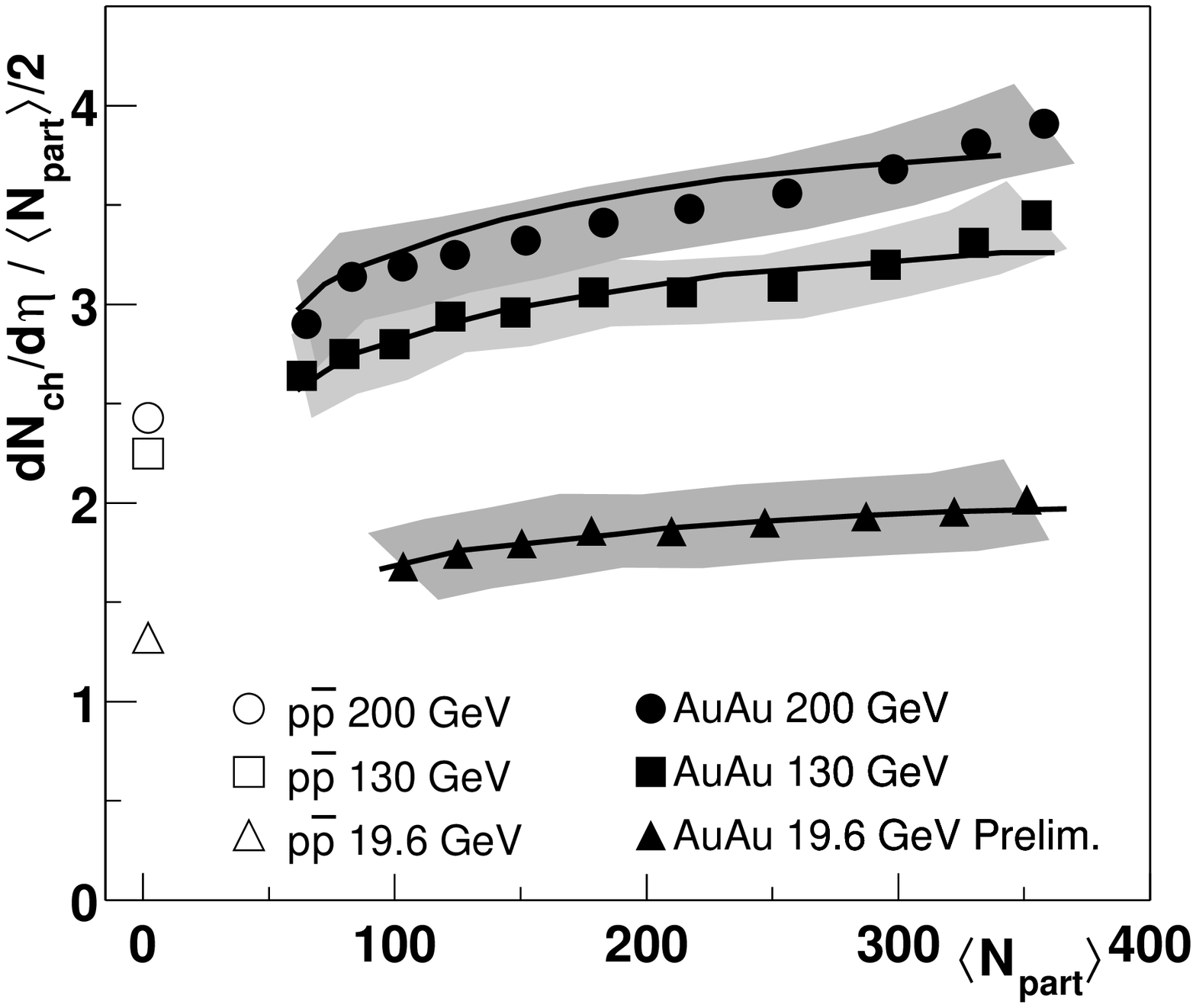,width=75mm}
\vspace*{-0.5cm}
\caption{Scaled pseudorapidity density at mid-rapidity,
$dN/d\eta/\langle N_{part}/2 \rangle$, as a function of
centrality for 19.6, 130, and 200~GeV. 
The curves correspond to predictions from the saturation 
model~\protect\cite{dima1,dima2}.}
\label{fig:Aneta}
\end{minipage}
\hspace{\fill}
%
\begin{minipage}[t]{75mm}
\epsfig{file=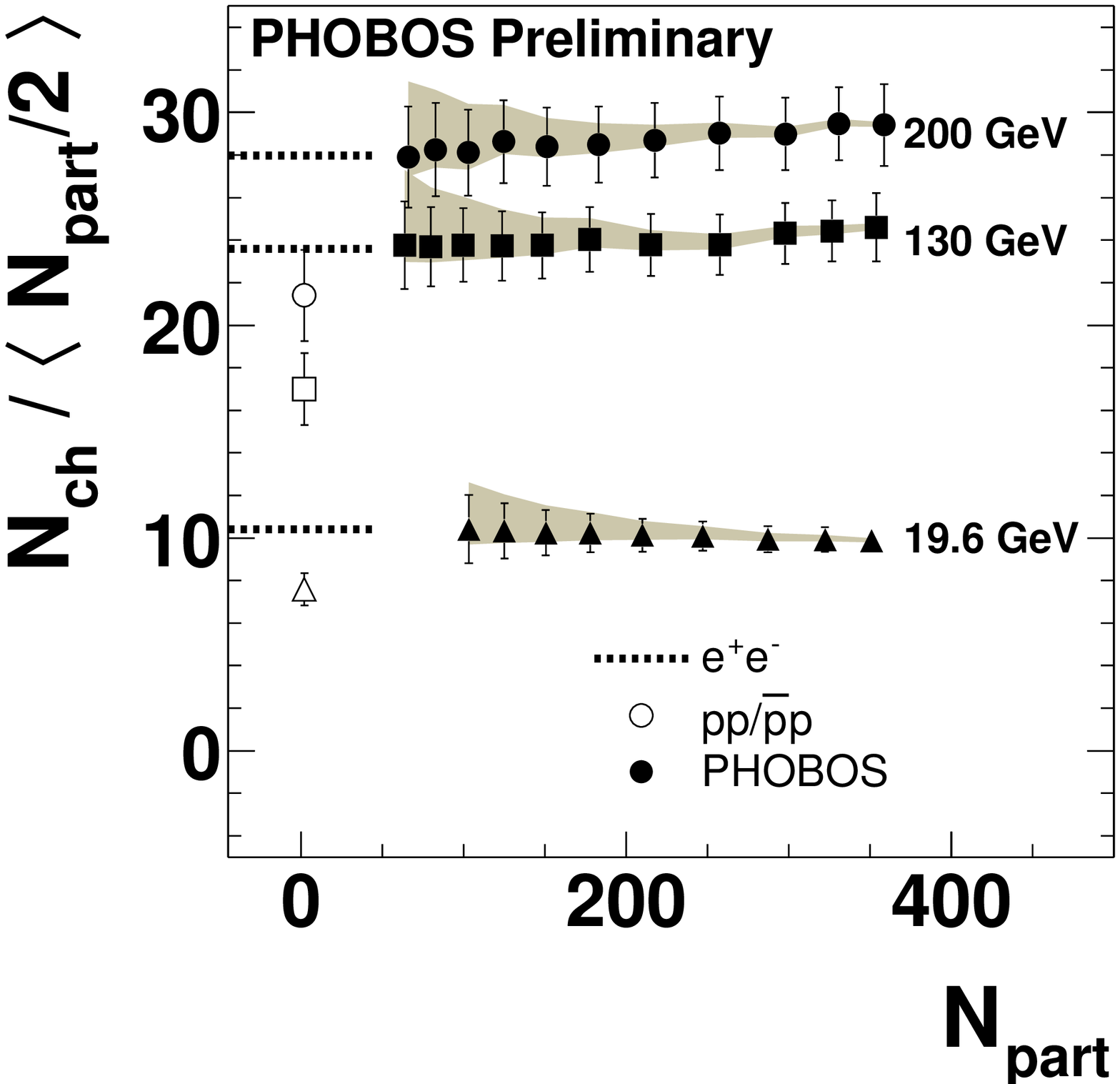,width=75mm}
\vspace*{-0.5cm}
\caption{Total charged multiplicity normalized by 
$\langle N_{part}/2 \rangle$ as a function of $N_{part}$ for AA.
Also shown are results for \ee\ and \ppbar\ data.}
\label{fig:Ntotcent}
\end{minipage}
\end{figure}

Participant scaling was first established in the context of the total
charged particle multiplicity produced in \pA\
collisions~\cite{Wounded}. In high energy AA collisions, small
deviations from participant scaling for mid-rapidity $dN/d\eta$ have
been seen. For instance, Figure~\ref{fig:Aneta} shows the mid-rapidity
$dN/d\eta$ scaled by $\avgNp/2$ measured with the same apparatus for
$\sqrtsnn$ = 19.6, 130, and 200~GeV. Results of this type are
sometimes interpreted in terms of a mixture of ``hard'' and ``soft''
components scaling as $N_{part}^{4/3}$ and $N_{part}$
respectively~\cite{dima1}. This interpretation is somewhat ambiguous,
however, since saturation models~\cite{dima1,dima2} describe the data
more economically, as can be seen in Figure~\ref{fig:Aneta}, and since
both dynamic and kinematic effects can shift particles around in
pseudorapidity as centrality changes.

Due to the large acceptance of the PHOBOS detector, we can integrate
the total charged particle production in these
collisions. Figure~\ref{fig:Ntotcent} shows $N_{ch}/\langle N_{part}/2
\rangle$ as a function of $\avgNp$ for the three different energies.
Also shown are the equivalent results for \pp\ and \ppbar\ as well as
\ee\ collisions at the same values of $\sqrt{s}$. It should be noted
that the \pp/\ppbar\ data are shown at their collision $\sqrt{s}$ for
this plot and not $\sqrt{s_{eff}}$. Three conclusions can be drawn
from this result. First, total charged multiplicity in high energy AA
collisions approximately scales with wounded nucleons
($N_{part}$). Second, total particle production in AA collisions per
participant pair is the same as the total particle production in \ee\
at the same energy for a broad range of centralities, not just for
central collisions. Finally, particle production in \pp\ and \ppbar\
collisions is reduced compared to AA and \ee.

While the overall charged particle production scales with $N_{part}$,
at high $\pT$, particle production is expected to scale with $N_{coll}
\propto N_{part}^{4/3}$. Deviations from this scaling at high $\pT$ are
likely to indicate high density effects in the initial or final state.
PHENIX and STAR have already shown that high $\pT$
hadron production is suppressed with respect to $N_{coll} \times
\sigma_{pp \rightarrow hX}$~\cite{jetq}. Recalling that
Figure~\ref{fig:Ntotcent} indicates that \pp\ collisions may not be the ideal
reference sample, we investigate the scaling of high $\pT$ hadron
production in AA collisions internally, using a mid-central \AuAu\
collision as a reference rather than \pp\ data.


Figure~\ref{fig:Christof} shows $d^2N/dyd\pT/\hlfNp$ as a function of
$\avgNp$ scaled by the value of the most peripheral point for
$0.5<y<1.3$ for various values of $\pT$.  The centrality bins used
here are specified in Figure~\ref{fig:spectra}.  The solid curve in
the lower right-hand panel is
$(N_{coll}/N_{part})/(N_{coll}/N_{part})_{periph.}$. This curve shows
the expectation if $N_{coll}$ scaling for
\begin{floatingfigure}[r]{8cm}
\vspace*{1.22em}
\epsfig{file=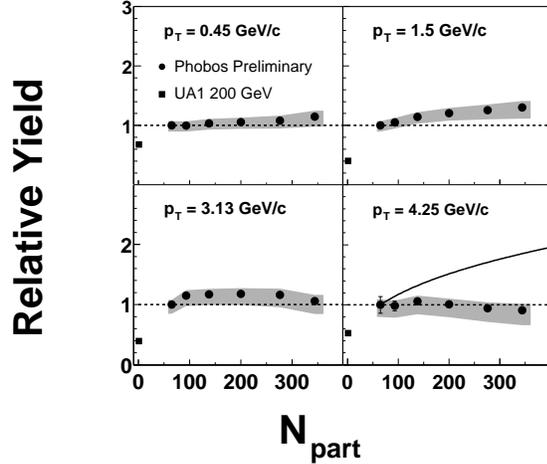,width=75mm}
\caption{Charged hadron yield in \AuAu\ per participant pair
($\avgNp/2$), normalized to the yield of the most peripheral bin
(45--50\%) as a function of $\avgNp$ for different $\pT$. 
The grey bands represent the
90\%~C.L.\ systematic error. The solid line in the lower right-hand
panel shows the expectation from $N_{coll}$ scaling.}
\label{fig:Christof}
\end{floatingfigure}
\noindent
\AuAu\ collisions held true
over the centrality range shown. This ratio varies by a factor of two
over this centrality range, and yet the particle production
approximately scales with $N_{part}$ for most $\pT$ bins. This
confirms the empirical observation that particle production at high
$\pT$ is suppressed with respect to the ``hard scaling''
expectation. However, we note the additional intriguing fact that, for
$N_{part}>65$, we see approximate $N_{part}$ scaling of both low and
high $\pT$ particle production. Mid-range particles with $\pT \sim
1.5$~GeV/c show a small violation of $N_{part}$ scaling, but are still
suppressed with respect to the na\"{\i}ve scaling expectations.  This
particular form of ``high $\pT$ suppression'' could be an indication
of initial state suppression (e.g. parton saturation) or that the
final state suppression (e.g. ``jet quenching'') reaches a geometric
maximum involving one power of length scale $R_{Au} \propto
N_{part}^{1/3}$. Of course, this apparent $N_{part}$ scaling could
also be accidental.

\section{SUMMARY}

The PHOBOS experiment has measured a systematic dataset of
Au+Au collisions at the RHIC collider. These data span an energy range
from $\sqrtsnn$ = 19.6 to 200 GeV, a pseudorapidity range from -5.4 to
+5.4 and a centrality range from 65 to 340 participating nucleons.  It
also includes particle spectra covering the $\pT$ range from 0.03 to
5.0~GeV/c. 

Three empirical observations have emerged from this dataset.  First,
there is clear evidence of limiting fragmentation in \AuAu\
collisions: an energy-independent region of $\eta'$ ($\eta' \equiv
\eta-y_{beam}$).  This energy-independent region grows with energy,
allowing only a limited region (if any) of longitudinal
boost-invariance.  Second, there is a striking, unexplained,
similarity of particle production in \ee\ and particle production per
participating nucleon pair in \AuAu\ collisions.  Finally, particle
production has been found to approximately scale with number of
participating nucleon pairs for $\avgNp>65$. This scaling occurs for
the total multiplicity and also for high $\pT$ particles (including
3~$<\pT<$~4.5~GeV/c).

These empirical observations serve, at least, to characterize heavy
ion collisions in an economical way and to challenge models. At best,
the observed scaling and universalities may point the way to a fairly
simple partonic description of this high density matter and thus
advance our knowledge of the strong interaction.

\vspace{1mm}
\noindent
{\small Acknowledgments: 
This work was partially supported by U.S. DOE grants
DE-AC02-98CH10886, DE-FG02-93ER40802, DE-FC02-94ER40818,
DE-FG02-94ER40865, DE-FG02-99ER41099, and W-31-109-ENG-38 as well as
NSF grants 9603486, 9722606 and 0072204.  The Polish group was
partially supported by KBN grant 2-P03B-10323.  The NCU group was
partially supported by NSC of Taiwan under contract NSC
89-2112-M-008-024.
}

\end{document}